\documentclass[a4paper]{article}

\usepackage{INTERSPEECH2019}
\usepackage{mathtools}
\usepackage{breqn}
\usepackage{tabularx}
\usepackage{tikz}
\usepackage{bbm}
\DeclareMathOperator*{\E}{\mathbb{E}}

\title{Enhancing audio quality for expressive Neural Text-to-Speech}
\name{Abdelhamid Ezzerg, Adam Gabrys, Bartosz Putrycz,  Daniel Korzekwa, Daniel Sáez-Trigueros, David McHardy, Kamil Pokora, Jakub Lachowicz, Jaime Lorenzo-Trueba, Viacheslav Klimkov}
\address{
  Amazon Text-to-Speech Research}
\email{ezzerg@amazon.co.uk}

\begin{document}

\maketitle
\begin{abstract}
  Artificial speech synthesis has made a great leap in terms of naturalness as recent Text-to-Speech (TTS) systems are capable of producing speech with similar quality to human recordings. However, not all speaking styles are easy to model: highly expressive voices are still challenging even to recent TTS architectures since there seems to be a trade-off between expressiveness in a generated audio and its signal quality. In this paper, we present a set of techniques that can be leveraged to enhance the signal quality of a highly-expressive voice without the use of additional data. The proposed techniques include: tuning the autoregressive loop's granularity during training; using Generative Adversarial Networks in acoustic modeling; and the use of Variational Auto-Encoders in both the acoustic model and the neural vocoder. We show that, when combined, these techniques greatly closed the gap in perceived naturalness between the baseline system and recordings by 39\% in terms of MUSHRA scores for an expressive celebrity voice.

\end{abstract}

\section{Introduction}
Artificial speech synthesis has seen a considerable change of paradigm: from the use of concatenative-based approaches \cite{Alan-W-Black,Qian2013AUT,Merritt2016DeepNN}, to leveraging modern Neural Text-to-Speech (NTTS) architectures such as Wavenet \cite{WaveNet} and Tacotron \cite{Tacotron}. Neural-based models are capable of synthesizing speech that rivals the real one in terms of quality while not being as constrained as concatenative methods in terms of phonetic coverage. Nonetheless, neural models are still data-hungry: training high-fidelity TTS systems using neural networks requires many hours of high-quality training data \cite{Latorre2018}.\\
In addition to the challenge of gathering high-quality training data, we observed a tradeoff between the level of expressiveness in a voice (measured using the variance of f0, energy and phonemes' durations within the training data) and the segmental quality of produced speech: while standard architectures were able to produce high quality speech for neutral voices, their produced speech for highly-expressive voices suffered from degradations in audio quality.\\
In this paper, we present techniques that we applied, on top of a standard architecture such as in \cite{Tacotron}, to enhance the speech quality of highly-expressive voice. 
The described techniques are:
\begin{itemize}
\item Increasing the degree of autoregression as the training progresses
\item  The use of adversarial training for improving the quality of generated spectrograms
\item The use of variational autoencoders (VAEs) with carefully selected latent representations at inference time
\item Training a neural vocoder conditioned on latent representations extracted using a pre-trained VAE
\end{itemize}

In Section 2 we will separately explain each of the applied techniques, Section 3 will present and discuss the result of applying the above-mentioned techniques on an expressive celebrity voice while Section 4 will be for conclusions.

\section{Proposed approach}

\subsection{Model's architecture}

\begin{figure}
  \centering
  \includegraphics[width=0.5\textwidth]{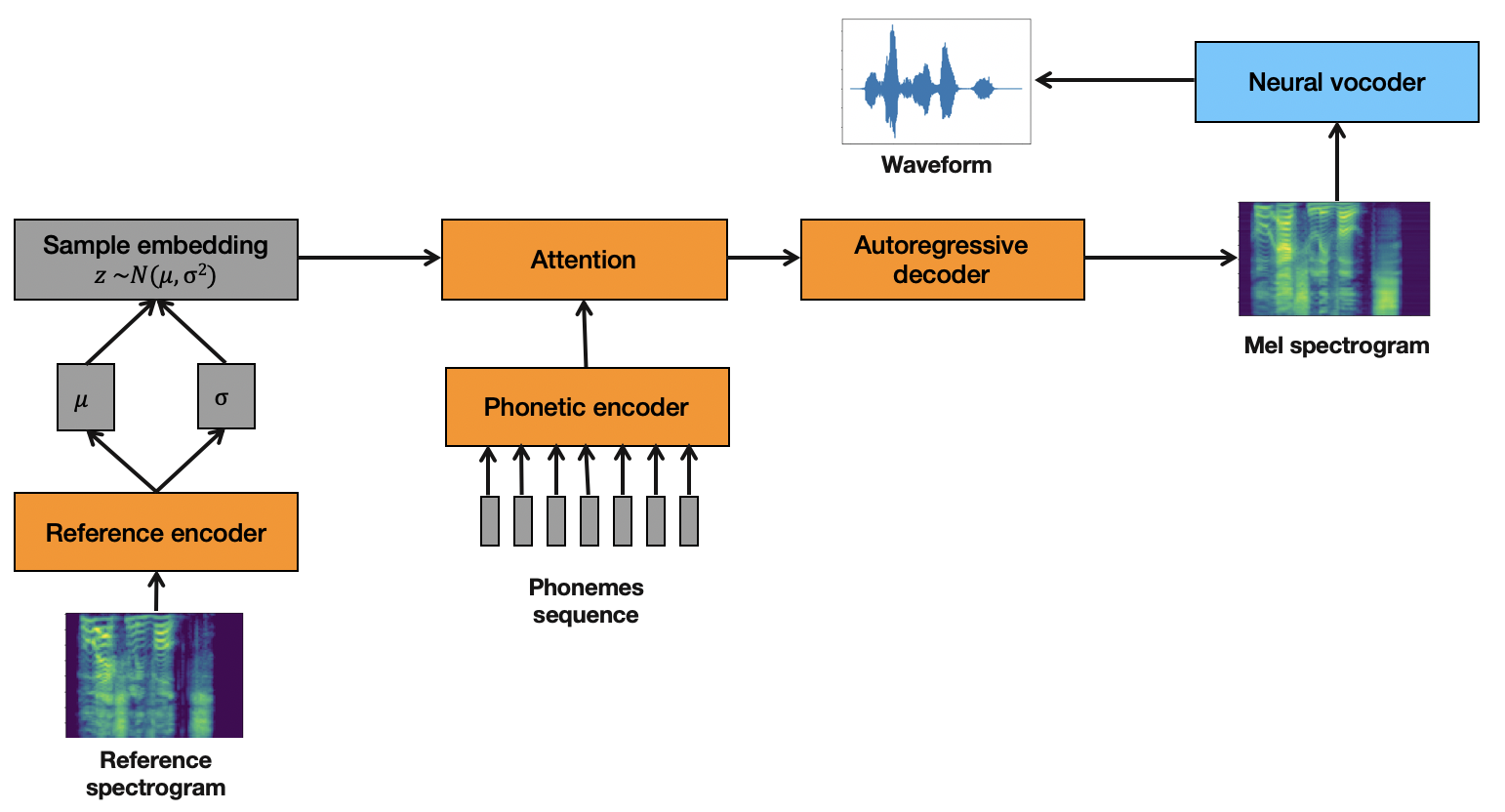}
  \caption{Overview of model architecture. The system can be broken into two parts: an acoustic model and a neural vocoder that produces waveform. Orange blocks highlight the building neural network blocks for the acoustic model while the neural vocoder is represented by a blue box.}

\end{figure}
The model we use comprises two main modules trained separately: an acoustic model which predicts a mel-spectrogram from an input sequence of phonemes, and a neural vocoder that predicts the waveform from the output of the acoustic model (see figure 1).\\
The acoustic model is a state-of-the-art sequence-to-sequence (seq2seq) neural network \cite{Tacotron,Prateek,Latorre2018,Shen2018} that leverages the attention mechanism \cite{Chorowski2015,Bahdanau2015}. The model was reinforced by the use of a Variational-Auto-Encoder (VAE) \cite{VAE2013} that takes the target mel-spectrogram as input and predicts the mean and variance of a Gaussian distribution from which a latent representation will be sampled. We use adversarial training \cite{GAN2014} in order to shift the distribution of predicted mel-spectrograms towards the distribution of target mel-spectrograms.The acoustic module models the following probability distribution:
\begin{equation}
  p(\textbf{y}_{1:M}) = \int\Pi_{m=1:M}p(\textbf{y}_m|\textbf{y}_{<m},\textbf{x}_{1:N}, \textbf{z})p(\textbf{z})d\textbf{z}
\end{equation}
Where $ \textbf{y} = \{\textbf{y}_1, \textbf{y}_2, ..., \textbf{y}_M\}$ is a sequence of mel-spectrogram frames, $ \textbf{x} = \{\textbf{x}_1, \textbf{x}_2, ..., \textbf{x}_N\}$ is a sequence of phoneme embeddings and $\textbf{z}$ is the VAE latent representation extracted from the target mel-spectrogram.\\ 
The vocoder is a parallel-Wavenet vocoder \cite{ParallelWavenet} with the addition of a VAE. The VAE-component takes as input the spectrogram predicted by the acoustic model and generates a latent representation (see section 2.5 for more details). The vocoder models the following distribution:
\begin{equation}
  p(\textbf{w}_{1:T}) = \Pi_{t=1:T}p(w_t|\mu(\textbf{s}_{<t},, \textbf{y}_{1:M}, \textbf{z}), \sigma(\textbf{s}_{<t},, \textbf{y}_{1:M}, \textbf{z}))
\end{equation}
Where $ \textbf{w} = \{w_1, w_2, ..., w_T\}$ is the waveform, $ \textbf{y} = \{\textbf{y}_1, \textbf{y}_2, ..., \textbf{y}_M\}$ is a sequence of mel-spectrogram frames, $\textbf{z}$ is the latent representation extracted from the target mel-spectrogram using the acoustic model's VAE module and $ \textbf{s} = \{\textbf{s}_1, \textbf{s}_2, ..., \textbf{s}_T\}$ is a sequence of noise sampled from a prior random variable that serves as input to the Inverse Autoregressive Flow (IAF) \cite{IAF2016} blocks of the neural vocoder.\\

\subsection{Tuning auto-regression levels}
The acoustic model predicting mel-spectrograms is an encoder-decoder architecture that uses location-sensitive attention mechanism \cite{Prateek,Chorowski2015}. The decoder is an LSTM-based autoregressive module: at each decoder step, the decoder predicts a set of mel-spectrogram frames based on frames predicted in its previous step. We observed that the interaction between the decoder and the attention mechanism led to instabilities when generating very long sequences. Such instability issues include mumbling and skipping over phonemes.\\ 
To alleviate the instability issues and help the convergence of the attention mechanism, we changed slightly the decoder's architecture enabling it to predict multiple spectrogram frames at a time instead of one. With multiple frames predicted per decoder step, the decoder needs less steps to produce the same output spectrogram, this reduction in number of steps helps prevents instabilities from accumulating during synthesis. This trick greatly improved the stability of the model; a finding that was also discussed in \cite{Tacotron}. However, the improved stability came at the cost of a decrease in segmental quality. In order to help stabilize the attention while maintaining the same level of audio quality, we tuned the decoder's number of outputs-per-step (ops) gradually from ops 5 to ops 2 within the same training. In order to tune the ops with no change to the architecture, we made the decoder predict the maximum ops (5 in this case) at all stages of the training. The decoder's output was then sliced depending on the current ops: for example, in ops 2 we would select only the first two predicted frames.\\ 
We made the decision to stop the tuning phase before reaching ops 1 (fully autoregressive model) because we observed, in the development process, that the ops 2 model generated samples of comparable quality to the ops 1 model while being faster.\\
Since the  autoregression's tuning approach attempts to improve on the instability issues on the decoder side, it can still be combined with orthogonal approaches that tackled the problem from the attention mechanism's side \cite{DCA2019, ForwardAttention2018,raffel2017online, SMA2019}.

\subsection{Using adversarial training for acoustic modeling}

Generative adversarial networks (GANs)\cite{goodfellow2014generative} have been successfully used to generate high quality images. \cite{styleGAN2018,larsen2015autoencoding,denton2015deep}. The adversarial loss incentivizes the generator to produce images that are indistinguishable from real ones, thus mitigating the over-smoothing effect observed when using traditional losses such as L1 or L2 alone \cite{GANLoss}. Following the progress made in image generation, GANs have started to be applied to TTS. Kaneko et al. \cite{Kaneko2017,Kaneko_2017_Interspeech} applied GANs to train the post-filter component of the acoustic model to produce sharper mel-spectrograms. Other GAN configurations were also explored, such as applying GANs on the waveform for speech enhancement \cite{pascual2017segan} and the use of GANs to mitigate exposure bias \cite{guo2019new}.\\
We explored the use of adversarial training to reduce over-smoothing in mel-spectrogram prediction via end-to-end training of our acoustic model. By adding an adversarial loss, we aim to encourage the network to output a spectrogram distribution that matches with that of the target and not only focus on the over-smoothing L1/ L2 losses. In our configuration (figure 2), the generator is the whole acoustic model and is trained using L1 loss between predicted and target mel-spectrograms plus the adversarial loss. The discriminator is trained to distinguish predicted spectrograms from target ones and is based on self-attention blocks following the same architecture as in \cite{SAGAN2018}. The following equations summarize the training losses of both the discriminator and the generator:

\begin{dmath}
  L_D=-\E_{x \sim p_{data}}{[min(0, -1-D(G(x)))]}-\E_{y \sim p_{data}}{[min(0, -1+D(y))]}
\end{dmath}

\begin{dmath}
L_G=\E_{x,y \sim p_{data}}{[\left\lVert G(x) -y \right\rVert_1]}+
    \alpha \E_{x,y \sim p_{data}}{[D(y)-D(G(x))]} -
     \beta(step) KLD(p_z, p_{prior})
\end{dmath}

Where D is the discriminator network, G is the generator network, x is the phoneme input sequence, y is the target mel-spectrogram sequence, $\alpha$ is a weighting factor used to balance the contributions of the adversarial loss and the L1 reconstruction loss, z is the latent representation sampled from the distribution $p_{z}$ whose parameters are predicted by the VAE encoder and $\beta$ is a weighting factor for the Kullback-Leibler Divergence (KLD) loss used for the VAE training (see section 2.4).

We used spectral normalization \cite{SpectralNormalization2018} on the discriminator side to stabilize the discriminator's training. We also observed that feeding the whole mel-spectrogram sequence to the discriminator gave worse results than feeding a small random window of mel-spectrogram frames. We think that this approach forced the discriminator to focus on short-term transitions in the audio, thus explaining the improved audio quality.

\begin{figure}
  \centering
  \includegraphics[width=0.45\textwidth]{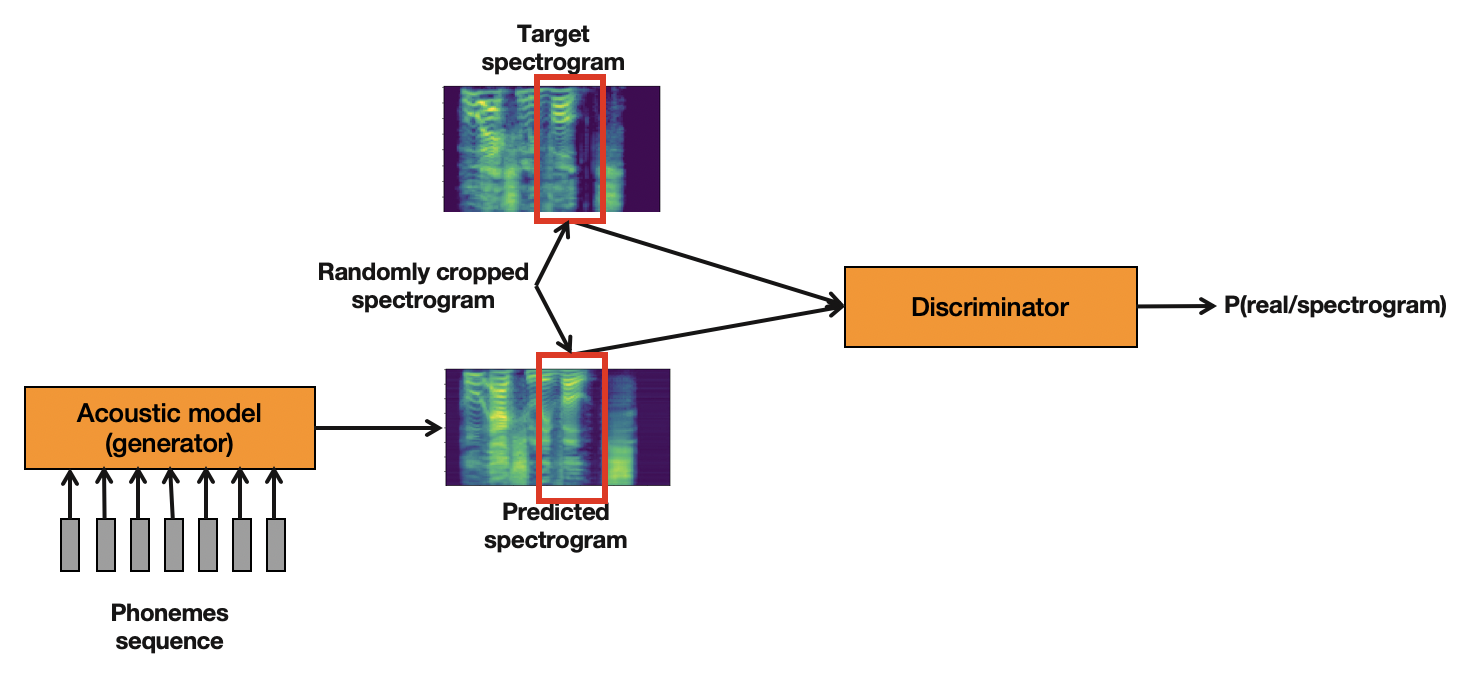}
  \caption{GAN training setup. The whole acoustic model is the generator. The discriminator network classifies its input spectrogram as real or predicted.}
\end{figure}

\subsection{Variational Auto-Encoders (VAEs)}
The acoustic model is conditioned on phonetic input which does not account for latent (i.e. prosodical) factors in the data. To be able to factorize these latent elements, we enhance the acoustic model via the addition of a Variational Auto-Encoder (VAE) \cite{VAE2013} which produces a latent representation predicted from the target spectrogram. Similar approaches were used for style modelling using continuous hierarchical embeddings \cite{hsu2018hierarchical} or discrete ones \cite{wang2018style}. The VAE module is a reference-encoder-like architecture made of a stack of convolutional neural networks followed by a BiLSTM and two projections that predict the mean and standard deviation of a 64-dimensional Gaussian distribution. The prior of the VAE latent vector is a Normal distribution with zero mean and unit variance. As such, the Kullback-Leibler Divergence (KLD) has a closed form equation.\\

When training with KLD loss, it is possible to observe KLD collapse: the decoder ignores the latent variable, thus keeping the posterior distribution similar to the uninformative standard Gaussian prior. To alleviate this issue, approaches such as annealing or introduction of skip connections have been proposed \cite{klCollapse1,klCollapse2, klCollapse3,klCollapse4}. We use a simple annealing scheme where the weighting factor of the KLD loss is gradually increased from 0 to 1 until a given step, after which the KLD loss is only periodically applied.\\ 

Another challenge faced while introducing VAE to our acoustic model is the selection of latent variable to use at inference time. Two main schemes can be used: sampling from the prior distribution of latent variables or providing a fixed latent representation at inference time for all utterances. For the second scheme, different variations can be used, such as using the mean of the prior distribution, using centroid (mean) computed over training data, or selecting a latent representation extracted from sampled utterances from the training set. We observed that the selected latent representation can have a big impact on the prosody and audio quality of generated samples. Furthermore, we observed that the latent representation extracted from spectrograms corresponding to utterances with flat/average prosody led to better observed segmental quality. After extensive listening, we chose a scheme where we use a latent vector extracted from an utterance with flat intonation for general speech, and a latent vector extracted from an utterance with rising intonation for yes/no questions. The acoustic model will use one of these latent representation at inference time depending on the domain.

\subsection{VAE-enhanced parallel Wavenet}
The vocoder is a Parallel Wavenet-like \cite{ParallelWavenet} architecture trained with probability density distillation and additional spectral loss term. To improve the vocoder's synthesized speech quality, we used a similar approach to \cite{Jonas2020} which conditioned both the teacher and student networks on additional VAE latent representation extracted from real-speech (section 2.4). Figure 3 shows how the VAE conditioning is performed.\\
The teacher model has a Wavenet-like architecture with mixture of 10 logistics where audio samples are conditioned on oracle mel-spectrograms and a 64-dimensional VAE latent representation extracted from the VAE encoder of the acoustic model. Mel-spectrogram frames are encoded by a 2-layers BiLSTM with 128 hidden size, they are then concatenated with a 64-dimensional VAE latent vector. We then apply an affine transformation, implemented as a 1x1 convolution, to the output concatenated vector. Finally, the conditioning representation is upsampled to align with audio samples. Every Residual Gated CNN  block uses 256-dimensional skip and gated channels. Filter activation is tanh and the gate activation is a sigmoid function.\\	
The student network  shares the same conditioning blocks with frozen weights with the teacher network. A logistic distribution is passed through a stack of 4 Inverse Autoregressive Flows \cite{IAF2016} with affine transform. The parameters of the affine transform are predicted by autoregressive conditioner blocks similar in architecture to Wavenet blocks \cite{WaveNet}. The flow
conditioners contain dilated convolutions with 10, 10, 10 and 30 layers. The last block uses the same dilation value growth and reset as the teacher network. The dilated convolutions use 64-dimensional gated channels with tanh filter activation and sigmoidal gate activation.\\
We tested the effect of the size of the VAE latent representation by comparing a 16-dimensional representation against a 64-dimensional one and we observed that, while the 16-dimension teacher had a higher audio quality, the student conditioned on 16-dimension VAE struggled to properly learn the teacher distribution and generated noisier samples. We also explored the use of VAE conditioning on the student only, given the already high audio quality of the teacher, and found out that the student network was unable to properly match the teacher's distribution. This behavior needs more investigation and may drive future work. Another detail to take into account is which latent representation to use at inference time. Two schemes were examined: the use of a centroid (mean) computed over the training data and computing the latent representation from the conditioning spectrogram. Extensive listening led us to the choice of using the centroid conditioning.

\begin{figure}
  \includegraphics[width=0.5\textwidth,height=5cm]{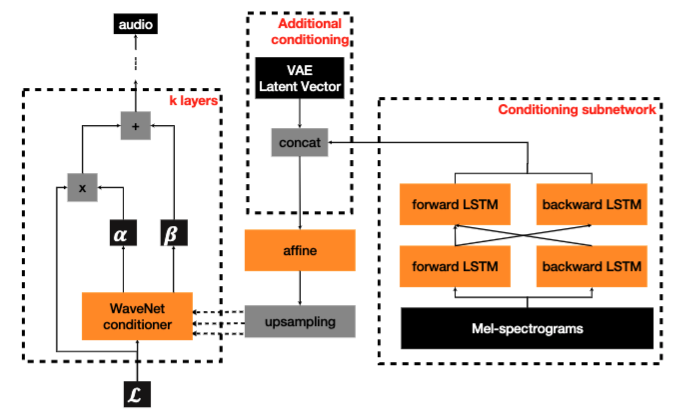}
  \caption{The architecture of the Parallel WaveNet-like neural vocoder. The additional conditioning block concatenates the mel-spectrogram conditioning with the latent representation extracted from the VAE reference encoder of the acoustic model.}
  
\end{figure}

\section{Results}

\subsection{Experiments}
To demonstrate the benefits of the techniques detailed in the previous section, we train two NTTS systems on highly-expressive data of a male, English-speaking, celebrity voice. The first system is a baseline model: an autoregressive acoustic model with 5 spectrogram frames given per each decoder step (with similar architecture to \cite{Prateek,Latorre2018}) followed by a parallel Wavenet vocoder. The second model is the baseline model enhanced by all four techniques described in Section 2 and we will refer to it as full-system in the remainder of the paper.\\ 
The acoustic model was trained separately from the vocoder using a batch size of 32. The training procedure for the auto-regression tuning was as follows: train the baseline with 5 output frames predicted per decoder step (denoted as ops for output-per-step), then tune the model as ops 4, followed by tuning the model as ops3 and finally as ops2. We used Adam optimizer with default parameters. Once the ops 2 tuning is finished, we tune the model using the introduced GAN module. In the GAN tuning phase, the $\beta_1$ parameter of the Adam optimizer was reduced. We train the model on 10 hours of expressive data of a male US voice.\\
In the vocoder's training procedure, we use a batch size of 16 and an Adam optimizer with default parameters. Learning rate decay wasn't used in student training, but it was used in the teacher network's training with a decay value of 0.95. We used Polyak averaging with a decay value of 0.999.\\
Our student was trained iteratively on 3 different saved snapshots of the teacher network, taken at different training steps, in order to to be able to train the student network while the teacher's training was still ongoing. We have observed through listening that the student trained this way produced better-sounding audios than the student trained only on the last snapshot of the teacher. This observation will need to be investigated, but we hypothesize that the teacher's distribution gets more complex and harder to model by the student in the teacher's later iterations and thus the iterative training provides the student with checkpoints that are easier to match during early training steps.\\

\subsection{Evaluation}
We conducted a MUSHRA (Multiple Stimuli with Hidden Reference and Anchor) evaluation test with three systems: the baseline, the full-improvements system and recordings from the speaker. The test was performed on a set of 160 utterances with varying lengths. Each utterance was evaluated by 15 native English speakers who were asked to rate the three systems based on naturalness. The results of the perceptual test can be viewed in figure 4. We report the following mean MUSHRA scores per system: baseline: 51.13, full-system: 64.04 and recordings: 84.04. These numbers translate into the full-system achieving a 39\% gap-closing between the baseline and the recordings.\\

\begin{figure}
  \includegraphics[width=0.5\textwidth,trim={0 0.15cm 0 0},clip]{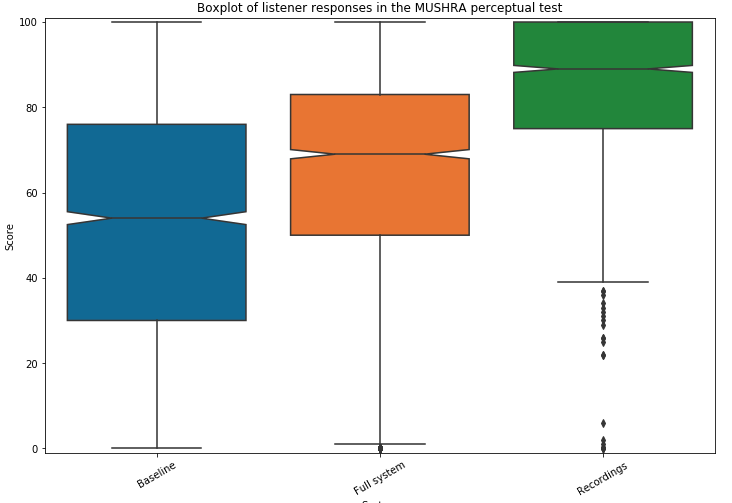}
  \caption{Boxplot of MUSHRA test's results. Systems from left to right: baseline, full-system and recordings. Mean score per system: baseline: 51.13, full-system: 64.04 and recordings: 84.04.}
  
\end{figure}

We also conducted VQA tests (Voice Quality Assessment) of the baseline and our proposed system where two native US English speakers were asked to report on issues they hear in audio files synthesized using both systems. The test set is comprised of 275 utterances with varying lengths and from different domains: questions, spelling, newscasting, etc. The issues were classified according to severity, from critical to almost unnoticeable (reported as minor). The reported issues covered audio quality, pronounciation issues and instabilities. Table 1 summarizes the VQA results, where we can observe a significant reduction in terms of reported issues. The biggest difference between the two systems was reported on audio buzziness issues: as the baseline system had 153 reported problems compared to 60 for the proposed system.\\

\begin{table}
  \begin{tabularx}{\linewidth}{|X|XXX|} 
   \hline
   System & Number of reported \textbf{critical} issues & Number of reported \textbf{medium} issues & Number of reported \textbf{minor} issues \\
   \hline
   Baseline  & 7 & 443 & 252 \\ 
   \hline
   Full-system & 0 & 231 & 123\\
   \hline
   \end{tabularx}
  \caption{Summary of VQA tests performed on both the baseline and the proposed system. It is worth noting that a single issue can be reported twice as there are two testers.}
\end{table}

In addition to the above MUSHRA test, we conducted an additional ablation test in order to rank the improvement made by each change. The test has the exact same setup as the previously described MUSHRA except that we use 4 systems instead of 3. The 4 systems are: full system without ops tuning, full system without GAN training, full system without VAE in acoustic model side and full system without VAE on neural vocoder's side. Table 2 summarizes the results of the MUSHRA.

\begin{table}
  \begin{tabularx}{\linewidth}{XX} 
   \hline
   System & MUSHRA score \\
   \hline
   w/o ops tuning  & 66.62 $\pm$ 1.06  \\ 
   \hline
   w/o GAN training & 67.64 $\pm$ 1.03 \\
   \hline
   w/o VAE in acoustic modeling  & 68.19 $\pm$ 1.03  \\
   \hline
   w/o VAE in vocoder  & 65.88 $\pm$ 1.07  \\
   \hline
   \end{tabularx}
  \caption{Mean MUSHRA score per system in ablation study.}
\end{table}
From the results of table 2, we observe that not all techniques are contributing equally to the overall improvements. Conditioning the Parallel Wavenet vocoder on VAE embeddings is the most impactful as the full system suffered the most without it. The second most important change is the tuning of ops within training. The GAN training and the VAE conditioning in the acoustic model side seem to have the least impact on the overall gains. This observation does not mean that the latter VAE should be discarded altogether since we noted, in section 2.2, that the component helped guide the question/non-question intonations.

\section{Conclusions}
The paper tackled the challenge of improving the audio quality of speech produced by a TTS system trained on highly-expressive data. To that end, we presented a compilation of techniques that ranged from acoustic modeling to vocoding. We then showed that, when combined, the proposed techniques improved the perceived quality which translated into a considerable increase in MUSHRA score and a reduction to the number of reported audio quality issues. We also run an ablation MUSHRA test to rank the impact of the proposed techniques.
%We presented a compilation of techniques that can be used to enhance the quality of audio produced by an NTTS system. The techniques are: tuning autoregression in the acoustic model during training, the introduction of VAE and GANs in acoustic modeling, and the conditioning of the Parallel Wavenet vocoder on latent vectors derived using the acoustic model's VAE. When combined, these techniques improved the perceived quality of the baseline system by a 39\% increase in relative MUSHRA scores and and cut the total number of reported audio quality and stability issues in half. Some of the proposed approaches, such as the use of GANs and the selection of VAE latent embedding to enhance the audio quality, still need further investigation, which could be the topic of future work.

%\section{Acknowledgements}

\bibliographystyle{IEEEtran}

\bibliography{mybib}

\begin{thebibliography}{10}
\providecommand{\url}[1]{#1}
\csname url@samestyle\endcsname
\providecommand{\newblock}{\relax}
\providecommand{\bibinfo}[2]{#2}
\providecommand{\BIBentrySTDinterwordspacing}{\spaceskip=0pt\relax}
\providecommand{\BIBentryALTinterwordstretchfactor}{4}
\providecommand{\BIBentryALTinterwordspacing}{\spaceskip=\fontdimen2\font plus
\BIBentryALTinterwordstretchfactor\fontdimen3\font minus
  \fontdimen4\font\relax}
\providecommand{\BIBforeignlanguage}[2]{{%
\expandafter\ifx\csname l@#1\endcsname\relax
\typeout{** WARNING: IEEEtran.bst: No hyphenation pattern has been}%
\typeout{** loaded for the language `#1'. Using the pattern for}%
\typeout{** the default language instead.}%
\else
\language=\csname l@#1\endcsname
\fi
#2}}
\providecommand{\BIBdecl}{\relax}
\BIBdecl

\bibitem{Alan-W-Black}
A.~W. Black and N.~Campbell., ``Optimising selection of units from speech
  databases for concatenative synthesis,'' 1995.

\bibitem{Qian2013AUT}
Y.~Qian, F.~K. Soong, and Z.-J. Yan, ``A unified trajectory tiling approach to
  high quality speech rendering,'' \emph{IEEE Transactions on Audio, Speech,
  and Language Processing}, vol.~21, pp. 280--290, 2013.

\bibitem{Merritt2016DeepNN}
T.~Merritt, R.~A.~J. Clark, Z.~Wu, J.~Yamagishi, and S.~King, ``Deep neural
  network-guided unit selection synthesis,'' \emph{2016 IEEE International
  Conference on Acoustics, Speech and Signal Processing (ICASSP)}, pp.
  5145--5149, 2016.

\bibitem{WaveNet}
\BIBentryALTinterwordspacing
A.~van~den Oord, S.~Dieleman, H.~Zen, K.~Simonyan, O.~Vinyals, A.~Graves,
  N.~Kalchbrenner, A.~W. Senior, and K.~Kavukcuoglu, ``Wavenet: {A} generative
  model for raw audio,'' \emph{CoRR}, vol. abs/1609.03499, 2016. [Online].
  Available: \url{http://arxiv.org/abs/1609.03499}
\BIBentrySTDinterwordspacing

\bibitem{Tacotron}
\BIBentryALTinterwordspacing
Y.~Wang, R.~J. Skerry{-}Ryan, D.~Stanton, Y.~Wu, R.~J. Weiss, N.~Jaitly,
  Z.~Yang, Y.~Xiao, Z.~Chen, S.~Bengio, Q.~V. Le, Y.~Agiomyrgiannakis,
  R.~Clark, and R.~A. Saurous, ``Tacotron: {A} fully end-to-end text-to-speech
  synthesis model,'' \emph{CoRR}, vol. abs/1703.10135, 2017. [Online].
  Available: \url{http://arxiv.org/abs/1703.10135}
\BIBentrySTDinterwordspacing

\bibitem{Latorre2018}
\BIBentryALTinterwordspacing
J.~Latorre, J.~Lachowicz, J.~Lorenzo{-}Trueba, T.~Merritt, T.~Drugman,
  S.~Ronanki, and K.~Viacheslav, ``Effect of data reduction on
  sequence-to-sequence neural {TTS},'' \emph{CoRR}, vol. abs/1811.06315, 2018.
  [Online]. Available: \url{http://arxiv.org/abs/1811.06315}
\BIBentrySTDinterwordspacing

\bibitem{Prateek}
\BIBentryALTinterwordspacing
N.~Prateek, M.~Lajszczak, R.~Barra{-}Chicote, T.~Drugman, J.~Lorenzo{-}Trueba,
  T.~Merritt, S.~Ronanki, and T.~Wood, ``In other news: {A} bi-style
  text-to-speech model for synthesizing newscaster voice with limited data,''
  \emph{CoRR}, vol. abs/1904.02790, 2019. [Online]. Available:
  \url{http://arxiv.org/abs/1904.02790}
\BIBentrySTDinterwordspacing

\bibitem{Shen2018}
\BIBentryALTinterwordspacing
J.~Shen, R.~Pang, R.~J. Weiss, M.~Schuster, N.~Jaitly, Z.~Yang, Z.~Chen,
  Y.~Zhang, Y.~Wang, R.~J. Skerry{-}Ryan, R.~A. Saurous, Y.~Agiomyrgiannakis,
  and Y.~Wu, ``Natural {TTS} synthesis by conditioning wavenet on mel
  spectrogram predictions,'' \emph{CoRR}, vol. abs/1712.05884, 2017. [Online].
  Available: \url{http://arxiv.org/abs/1712.05884}
\BIBentrySTDinterwordspacing

\bibitem{Chorowski2015}
\BIBentryALTinterwordspacing
J.~Chorowski, D.~Bahdanau, D.~Serdyuk, K.~Cho, and Y.~Bengio, ``Attention-based
  models for speech recognition,'' \emph{CoRR}, vol. abs/1506.07503, 2015.
  [Online]. Available: \url{http://arxiv.org/abs/1506.07503}
\BIBentrySTDinterwordspacing

\bibitem{Bahdanau2015}
\BIBentryALTinterwordspacing
D.~Bahdanau, K.~Cho, and Y.~Bengio, ``Neural machine translation by jointly
  learning to align and translate,'' \emph{ICLR}, 2015. [Online]. Available:
  \url{https://arxiv.org/abs/1409.0473}
\BIBentrySTDinterwordspacing

\bibitem{VAE2013}
\BIBentryALTinterwordspacing
D.~P. Kingma and M.~Welling, ``Auto-encoding variational bayes,'' \emph{arXiv
  preprints}, 2013. [Online]. Available: \url{https://arxiv.org/abs/1312.6114}
\BIBentrySTDinterwordspacing

\bibitem{GAN2014}
\BIBentryALTinterwordspacing
I.~J. Goodfellow, J.~Pouget-Abadie, M.~Mirza, B.~Xu, D.~Warde-Farley, S.~Ozair,
  A.~Courville, and Y.~Bengio, ``Generative adversarial networks,'' \emph{arXiv
  preprints}, 2014. [Online]. Available: \url{https://arxiv.org/abs/1406.2661}
\BIBentrySTDinterwordspacing

\bibitem{ParallelWavenet}
\BIBentryALTinterwordspacing
A.~van~den Oord, Y.~Li, I.~Babuschkin, K.~Simonyan, O.~Vinyals, K.~Kavukcuoglu,
  G.~van~den Driessche, E.~Lockhart, L.~C. Cobo, F.~Stimberg, N.~Casagrande,
  D.~Grewe, S.~Noury, S.~Dieleman, E.~Elsen, N.~Kalchbrenner, H.~Zen,
  A.~Graves, H.~King, T.~Walters, D.~Belov, and D.~Hassabis, ``Parallel
  wavenet: Fast high-fidelity speech synthesis,'' \emph{CoRR}, vol.
  abs/1711.10433, 2017. [Online]. Available:
  \url{http://arxiv.org/abs/1711.10433}
\BIBentrySTDinterwordspacing

\bibitem{IAF2016}
\BIBentryALTinterwordspacing
D.~P. Kingma, T.~Salimans, and M.~Welling, ``Improving variational inference
  with inverse autoregressive flow,'' \emph{CoRR}, vol. abs/1606.04934, 2016.
  [Online]. Available: \url{http://arxiv.org/abs/1606.04934}
\BIBentrySTDinterwordspacing

\bibitem{DCA2019}
E.~Battenberg, R.~Skerry-Ryan, S.~Mariooryad, D.~Stanton, D.~Kao, M.~Shannon,
  and T.~Bagby, ``Location-relative attention mechanisms for robust long-form
  speech synthesis,'' 2019.

\bibitem{ForwardAttention2018}
\BIBentryALTinterwordspacing
J.-X. Zhang, Z.-H. Ling, and L.-R. Dai, ``Forward attention in sequence-
  to-sequence acoustic modeling for speech synthesis,'' \emph{2018 IEEE
  International Conference on Acoustics, Speech and Signal Processing
  (ICASSP)}, Apr 2018. [Online]. Available:
  \url{http://dx.doi.org/10.1109/ICASSP.2018.8462020}
\BIBentrySTDinterwordspacing

\bibitem{raffel2017online}
C.~Raffel, M.-T. Luong, P.~J. Liu, R.~J. Weiss, and D.~Eck, ``Online and
  linear-time attention by enforcing monotonic alignments,'' 2017.

\bibitem{SMA2019}
M.~He, Y.~Deng, and L.~He, ``Robust sequence-to-sequence acoustic modeling with
  stepwise monotonic attention for neural tts,'' 2019.

\bibitem{goodfellow2014generative}
I.~J. Goodfellow, J.~Pouget-Abadie, M.~Mirza, B.~Xu, D.~Warde-Farley, S.~Ozair,
  A.~Courville, and Y.~Bengio, ``Generative adversarial networks,'' 2014.

\bibitem{styleGAN2018}
T.~Karras, S.~Laine, and T.~Aila, ``A style-based generator architecture for
  generative adversarial networks,'' 2018.

\bibitem{larsen2015autoencoding}
A.~B.~L. Larsen, S.~K. Sønderby, H.~Larochelle, and O.~Winther, ``Autoencoding
  beyond pixels using a learned similarity metric,'' 2015.

\bibitem{denton2015deep}
E.~Denton, S.~Chintala, A.~Szlam, and R.~Fergus, ``Deep generative image models
  using a laplacian pyramid of adversarial networks,'' 2015.

\bibitem{GANLoss}
\BIBentryALTinterwordspacing
W.~Lotter, G.~Kreiman, and D.~D. Cox, ``Unsupervised learning of visual
  structure using predictive generative networks,'' \emph{CoRR}, vol.
  abs/1511.06380, 2015. [Online]. Available:
  \url{http://arxiv.org/abs/1511.06380}
\BIBentrySTDinterwordspacing

\bibitem{Kaneko2017}
T.~{Kaneko}, H.~{Kameoka}, N.~{Hojo}, Y.~{Ijima}, K.~{Hiramatsu}, and
  K.~{Kashino}, ``Generative adversarial network-based postfilter for
  statistical parametric speech synthesis,'' in \emph{2017 IEEE International
  Conference on Acoustics, Speech and Signal Processing (ICASSP)}, March 2017,
  pp. 4910--4914.

\bibitem{Kaneko_2017_Interspeech}
T.~Kaneko, S.~Takaki, H.~Kameoka, and J.~Yamagishi, ``Generative adversarial
  network-based postfilter for stft spectrograms,'' in \emph{The Annual
  Conference of the International Speech Communication Association
  (Interspeech)}, August 2017.

\bibitem{pascual2017segan}
S.~Pascual, A.~Bonafonte, and J.~Serrà, ``Segan: Speech enhancement generative
  adversarial network,'' 2017.

\bibitem{guo2019new}
H.~Guo, F.~K. Soong, L.~He, and L.~Xie, ``A new gan-based end-to-end tts
  training algorithm,'' 2019.

\bibitem{SAGAN2018}
H.~Zhang, I.~Goodfellow, D.~Metaxas, and A.~Odena, ``Self-attention generative
  adversarial networks,'' 2018.

\bibitem{SpectralNormalization2018}
T.~Miyato, T.~Kataoka, M.~Koyama, and Y.~Yoshida, ``Spectral normalization for
  generative adversarial networks,'' 2018.

\bibitem{hsu2018hierarchical}
W.-N. Hsu, Y.~Zhang, R.~J. Weiss, H.~Zen, Y.~Wu, Y.~Wang, Y.~Cao, Y.~Jia,
  Z.~Chen, J.~Shen, P.~Nguyen, and R.~Pang, ``Hierarchical generative modeling
  for controllable speech synthesis,'' 2018.

\bibitem{wang2018style}
Y.~Wang, D.~Stanton, Y.~Zhang, R.~Skerry-Ryan, E.~Battenberg, J.~Shor, Y.~Xiao,
  F.~Ren, Y.~Jia, and R.~A. Saurous, ``Style tokens: Unsupervised style
  modeling, control and transfer in end-to-end speech synthesis,'' 2018.

\bibitem{klCollapse1}
Z.~Yang, Z.~Hu, R.~Salakhutdinov, and T.~Berg-Kirkpatrick, ``Improved
  variational autoencoders for text modeling using dilated convolutions,''
  2017.

\bibitem{klCollapse2}
A.~B. Dieng, Y.~Kim, A.~M. Rush, and D.~M. Blei, ``Avoiding latent variable
  collapse with generative skip models,'' 2018.

\bibitem{klCollapse3}
Y.~Kim, S.~Wiseman, A.~C. Miller, D.~Sontag, and A.~M. Rush, ``Semi-amortized
  variational autoencoders,'' 2018.

\bibitem{klCollapse4}
T.~Zhao, R.~Zhao, and M.~Eskenazi, ``Learning discourse-level diversity for
  neural dialog models using conditional variational autoencoders,'' 2017.

\bibitem{Jonas2020}
J.~Rohnke, T.~Merritt, J.~Lorenzo-Trueba, A.~Gabrys, V.~Aggarwal, A.~Moinet,
  and R.~Barra-Chicote, ``Parallel wavenet conditioned on vae latent vectors,''
  2020.

\end{thebibliography}

\end{document}